\documentclass{article}
\usepackage{arxiv}
\usepackage[utf8]{inputenc} 
\usepackage[T1]{fontenc}    
\usepackage[breaklinks=true,colorlinks=true,linkcolor=blue,urlcolor=blue,citecolor=blue]{hyperref}
\usepackage{url}            
\usepackage{booktabs}       
\usepackage{amsfonts} 
\usepackage{physics}      
\usepackage{nicefrac}       
\usepackage{microtype}      
\usepackage{lipsum}
\usepackage{graphicx}
\usepackage{upgreek}
\usepackage{cancel}
\usepackage{soul}
\usepackage{mathptmx}
\usepackage{amssymb}
\title{Robust Polarization Gradient Cooling of Trapped Ions}

\author{
\\Wenbing Li\,$^{1,2*}$, Sebastian Wolf\,$^{1*}$,  Lukas Klein\,$^{1}$, Dmitry Budker\,$^{1,2,3,4}$,\\
Christoph E. Düllmann\,$^{2,3,5,6}$ and Ferdinand Schmidt-Kaler\,$^{1,2}$\\
\\
 $^1$\,QUANTUM, Institut für Physik, Johannes Gutenberg-Universität Mainz, 55128 Mainz, Germany\\
 $^2$\,Helmholtz-Institut Mainz, 55099 Mainz, Germany\\
 $^3$\,PRISMA Cluster of Excellence, Johannes Gutenberg-Universit\"at Mainz, 55128 Mainz, Germany\\
 $^4$\,Department of Physics, University of California, Berkeley, CA 94720-7300, USA\\
 $^5$\,Department Chemie, Johannes Gutenberg-Universit\"at Mainz, 55128 Mainz, Germany\\
 $^6$\,GSI Helmholtzzentrum für Schwerionenforschung GmbH, 64291 Darmstadt, Germany\\
  \\
 $^*$\,Authors to whom any correspondence should be addressed\\
\texttt{wenbingli@uni-mainz.de} \\
\texttt{wolfs@uni-mainz.de}
}

\begin{document}
\maketitle
\begin{abstract}
We implement three-dimensional polarization gradient cooling (PGC) of trapped ions.
Counter-propagating laser beams near $393\,$nm impinge in lin$\,\perp\,$lin configuration, at a frequency below the S$_{1/2}$ to P$_{3/2}$ resonance in $^{40}$Ca$^+$. 
Our measurements demonstrate that cooling with laser beams detuned to lower frequencies from the resonance is robust against an elevated phonon occupation number and works continuously in the crossover from regular Doppler cooling to detunings of tens of linewidths. It is thus robust against heating events and also works well for an initial ion motion far out of the Lamb-Dicke regime.
We show that PGC performance strongly depends on residual micromotion and find PGC working for a micromotion modulation index $\beta\leq 0.1$.
Still, we find that the spectral impurity of the laser field affects both, cooling rates and cooling limits. Thus, a Fabry-P\'{e}rot cavity filter is employed to efficiently suppress amplified spontaneous emission of the diode laser. 
We demonstrate mean phonon numbers for a single ion of $5.4(4)$ at a trap frequency of $2\pi \times 285\,$kHz and $3.3(4)$ at $2\pi\times480\,$kHz, in the axial and radial directions, respectively.
\end{abstract}


\section{Introduction} \label{Sec:Introduction}

Laser-cooled trapped ions are widely used in quantum simulation~\cite{Zhang17,Blatt2012,Johanning2009} and quantum computation~\cite{Janine2021,Blatt2008,Cirac1995}, for high-performance optical clocks~\cite{Brewer2019,Keller2019,Huang2016,Chou2010} or precision spectroscopy~\cite{Pyka2013,ulm2013observation,Roos2006,Nagerl2010,Clark2010}. For those applications, Coulomb crystals are formed under Doppler cooling~\cite{Phillips1998,Chalony2011,Eeschner2003}. However, in order to improve the fidelity of quantum gate operations in an ion-based quantum computer, to reduce systematic errors in an optical ion clock, and in precision spectroscopy applications, one may require deeper cooling of the ion degrees of motion, including collective modes in the ion crystal.

Nearly perfect ground state cooling is routinely achieved by resolved sideband (SB) cooling techniques which either employ a narrow dipole-forbidden transition~\cite{Diedrich1989,Poulsen2012,Goodwin2016} or Raman transitions between hyperfine or Zeeman ground state levels~\cite{Che2017,Seck2016,Thompson2013}. For a crystal of $N$ ions, the number of collective modes is $3N$ such that sequentially cooling on all resolved sidebands becomes increasingly impractical for large $N$. However, the spectral lineshape of a three-level system may be tailored by electromagnetically induced transparency (EIT) such that multiple collective modes are cooled simultaneously~\cite{Feng2020,Roos2000,Lechner2016,Qiao2021}. This bandwidth of cooling - well suited to cool a collection of collective modes - may be further increased in the polarization gradient cooling (PGC) method. In the pioneering theoretical work by C.\,Cohen-Tannoudji and co-workers~\cite{Claude1990,Cohen1998}, sub-Doppler PGC experimental results on atomic ensembles obtained by S.\,Chu, W.\,D.\,Phillips et al. were explained, for which all three were awarded the Physics Nobel prize in 1997~\cite{Wineland1992,Chu1991,Dalibard1989}. Also for trapped ions, sub-Doppler cooling was proposed~\cite{Cirac1993,Yoo1993} and led to first experiments by PGC~\cite{Birkl1994}. Recently, PGC was demonstrated for a single trapped ion in three dimensions, and for small linear crystals~\cite{ejtemaee2017}, highlighting the large cooling bandwidth of this method. PGC was extended to linear ion crystals with up to 51 ions and 2D-crystals in zigzag configuration with 22 ions along a single trap axis~\cite{Joshi2020}. 

The most appealing application of PGC is sub-Doppler multi-mode cooling over a large band of motional frequencies. The PGC cooling bandwidth exceeds that of EIT cooling, leading to a fast reduction of phonon numbers even for a large number of vastly different mode frequencies, significantly below Doppler cooling limits. In this work, we implement 3D PGC with red detuning (the laser frequency below resonance of the dipole transition S$_{1/2}\leftrightarrow$\,P$_{3/2}$ near $393\,$nm in singly charged $^{40}$Ca$^+$). This technique demonstrates robust sub-Doppler cooling for trapped ions far beyond Lamb-Dicke regime (LDR). We investigate, for the first time, the residual micromotion is a significant effect on PGC, which is only accessible when employing beams with a nonvanishing projection on the radial direction of the Paul trap. We find that the red detuning allows us to implement PGC even with excessively high initial phonon numbers, orders of magnitude above that which had been reported so far. In this configuration, using red detuned light fields, PGC does not require Doppler precooling anymore. In addition, PGC with red detuned light allows us to study the crossover between PGC and Doppler cooling, e.g. to optimize the cooling rate. With small frequency detuning, the parasitic Doppler heating of the lights for PGC that also can work for Doppler cooling is reduced during PGC, which is especially important for fast PGC with light detuned by only a few linewidths of $2\pi\times23\,$MHz from the dipole transition. Such robust PGC may be advantageous in case of non-ideal Doppler precooling or if the ions are exposed to strong heating events. We see an application case for externally generated (exotic) ions, injected into a Paul trap, captured and then sympathetically cooled down and incorporated into a laser-cooled host ion crystal. Specifically, we aim for precision spectroscopy of injected ions of various thorium isotopes and the low-energy isomer in $^{229}$Th~\cite{Berning2019,Stopp2019,Hass2020,Seiferle2019}.

The paper is organized as follows: We start describing the PGC scheme in Sec.\,\ref{Levels and transitons for PGC} and the ion trap and laser experimental setup in Sec.\,\ref{experimental setup}. And then present the PGC cooling results for a single ion and for a 4-ion linear crystal. Furthermore, we study the robustness of PGC in typically encountered experimental situations such as excessively high initial phonon occupation number, an imperfect localization of the ion position at the trap center. in Sec.\,\ref{experimantal characterization PGC}. We investigate the laser sources for PGC with spectral impurity in Sec.\,\ref{techniquee for improving PGC}. This results demonstrate PGC as a versatile method for preparing ions in low axial and radial vibrational states. 


\section{Levels and transitions for polarization gradient cooling in $^{40}$Ca$^+$}\label{Levels and transitons for PGC}

\begin{figure}[t!]
\centering
\includegraphics[width=1\textwidth]{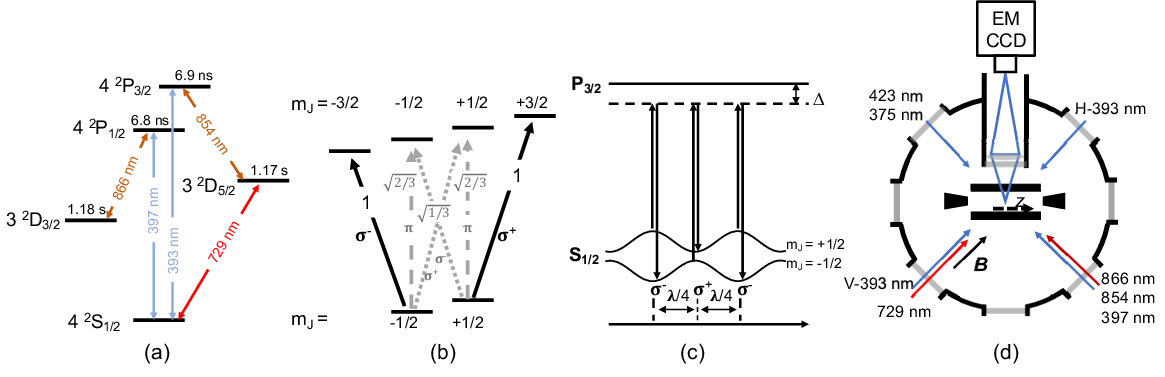}
\caption{(a) Levels and relevant transitions and lifetimes in $^{40}$Ca$^+$. (b) Energy levels and transitions involved in PGC. A static magnetic field yields a Zeeman splitting of $2\pi\times9.343\,$MHz in the ground S$_{1/2}$ state. The Clebsch-Gordan coefficients are shown for the transitions indicated with arrows. (c) Spatially varying light polarization results in position-dependent light shifts. The frequency of the PGC laser beams near 393\,nm is detuned by $\Delta$ from the S$_{1/2}\leftrightarrow\,$P$_{3/2}$ transition, and the wave vector $\mathbf{k}$ of both counter-propagating beams is aligned parallel to the magnetic field. (d) Geometry of trap, laser beams and magnetic field $\mathbf{B}$. The counter-propagating PGC beams are polarized horizontally (H-) and vertically (V-). Laser beams near $423\,$nm and $375\,$nm serve for photoionization loading, laser beams near $397\,$nm, $866\,$nm, $854\,$nm are used for Doppler cooling, and the ion fluorescence is imaged on a EMCCD camera. Light tuned near $729\,$nm is employed for resolved sideband spectroscopy.}
\label{fig:2LevelandLights} 
\end{figure}

Levels and relevant transitions in $^{40}$Ca$^+$ are shown in Fig.\,\ref{fig:2LevelandLights}\,(a) and the most important transition for PGC near 393~nm with its Zeeman components in Fig.\,\ref{fig:2LevelandLights}\,(b). The dipole-allowed transition from S$_{1/2}$ to P$_{3/2}$ is excited off-resonantly by a pair of counter-propagating $393\,$nm beams in lin$\,\perp\,$lin configuration, forming thus a periodically varying spatial polarization gradient lattice along this direction with alternating $\sigma^+, \pi, \sigma^-, \pi$, ... etc. polarization. If the frequency is red detuned relative to the S$_{1/2}$ to P$_{3/2}$ transition by $\Delta$, the variation of the Clebsch-Gordan coefficients leads to a periodically varying ac-Stark shift of the ground state Zeeman sublevels~\cite{Wineland1992,Dalibard1989}, as shown in Fig.\,\ref{fig:2LevelandLights}\,(c). If the ion moves along the direction of this lattice, predominantly either  $m_{J}=+1/2\,\leftrightarrow\,m_{J}=+3/2$ or  $m_{J}=-1/2\,\leftrightarrow\,m_{J}=-3/2$  transitions are alternately excited, in spatial regions with polarization of either $ \sigma^{+}$ or $ \sigma^{-}$, respectively. From the combination of ac-light shift and optical pumping, the ion will lose kinetic energy when climbing a potential hill. The wave vectors of the PGC beams are parallel with the magnetic field $B$. Its projections with the trap directions $(x, y, z)$ have angles ($60^\circ$, $60^\circ$, $45^\circ$) as shown in Fig.\,\ref{fig:2LevelandLights}\,(d). Thus, in this setting PGC is acting in all three dimensions, and cools axial and radial modes of vibration. The magnetic field strength of $\approx$ 343~$\mu$T is supplied with a set of permanent magnets which defines the quantization axis and leads to a splitting of $2\pi\times9.343\,$MHz between the two Zeeman sublevels of the S$_{1/2} $ manifold. 

Calcium atoms evaporated from an oven are photoionized resonantly in two photons step scheme with beams at wavelength of $423\,$nm and $375\,$nm and captured in the trap. For Doppler cooling, $397\,$nm light is red detuned from the dipole-allowed S$_{1/2}$ to P$_{1/2}$ transition by $\Delta \simeq -0.5 \Gamma$. Pumping out of the metastable D states is accomplished with beams near $866\,$nm and $854\,$nm. A beam near $729\,$nm is aligned parallel to the magnetic field direction and used for resolved sideband spectroscopy driving the narrow quadrupole $\ket{\text{S}_{1/2}, m_J=1/2}\, \leftrightarrow \ket{\text{D}_{5/2}, m_J=3/2}$ transition for revealing the ion phonon mode occupation numbers. 

\section{Experimental setup and procedures}\label{experimental setup}

We designed and built a linear Paul trap with blade-shaped RF (radio frequency) and DC (static voltage) electrodes which are integrally machined out of one piece of stainless-steel. The RF electrodes feature a distance of $2\,$mm from the trap center. We operate the trap at RF frequency $\Omega=2\pi\times5.83\,$MHz with a peak--to--peak amplitude of 550\,V to the RF pair of electrodes and generate trap frequencies of ($\omega_{x}, \omega_{y}) = 2\pi\times(483, 480)\,$kHz for the two radial directions. The DC pair is grounded, or kept at a small (<$10\,$V) voltage lifting further the degeneracy between radial trap frequencies. The axial confinement is provided by two end-cap electrodes at $14\,$mm distance. For a DC voltage of $+800\,$V, an axial frequency $\omega_z=2\pi \times 220\,$kHz is created along the z-direction of the trap.

The laser system used in the experiment includes external-cavity diode lasers\footnote{All lasers are TOPTICA DL pro.} operating near $397\,$nm for Doppler cooling, $866\,$nm and $854\,$nm for repumping, and $393\,$nm for PGC lattice. In addition, $729\,$nm light for probing the narrow quadrupole transition is supplied with a DL TA pro\,(Tapered Amplifier) locked to a high finesse optical cavity via Pound-Drever-Hall\,(PDH) resulting in a short-term frequency instability of $\leq$100\,Hz. The laser beams are switched by acousto-optic modulators\,(AOMs) operated in double pass configuration. All laser beams are coupled to the setup via polarization-maintaining single-mode fibers to ensure alignment stability. Laser-induced ion fluorescence is collected with a lens and imaged at a magnification of $\approx$11.3 onto an electron multiplying charge-coupled device\,(EMCCD) camera.

We investigate PGC with the following experimental sequence. First, the Ca ions are Doppler cooled with $397\,$nm laser light for $5\,$ms. The repumping light at $866\,$nm is switched on during Doppler cooling to prevent accumulation of ions in the D$_{3/2}$ state. Second, the PGC laser beams at $393\,$nm are applied to the ions at orthogonal linear polarization and along the magnetic field. The two beams have a frequency difference of $50\,$kHz to create a moving polarization gradient. This is needed so that all ion wave packets experience a significant polarization gradient lattice at some point during the PGC period. This should not be necessary for ion wave packets that are larger than a period of the gradient; our experimental situation corresponds to the borderline of this regime. We started with a value that is smaller than the axial trap frequency. After that, we tried to optimize the value by checking the final phonon number with different frequencies. However, we did not find that there was significant dependence even down to zero frequency difference. We conjecture that this is due the inherent phase fluctuations on both light paths of length $\sim 1.2\,$m. The intensity of the two $393\,$nm laser beams are calibrated with the fluorescence intensity detected with the EMCCD camera. During a PGC period, the light at $866\,$nm and at $854\,$nm is switched on to pump the populations out of the D$_{3/2}$ and D$_{5/2}$ states. Third, the ions are prepared in the $\ket{\text{S}_{1/2}, m_J=1/2}$ state via frequency-resolved optical pumping on the $\ket{\text{S}_{1/2},m_J=-1/2}\, \leftrightarrow\, \ket{\text{D}_{5/2}, m_J=1/2}$ transition by a laser pulse at $729\,$nm combined with light at $854\,$nm. Finally, the motional state is probed by addressing the carrier and first-order sideband of the  $\ket{\text{S}_{1/2}, m_J=1/2}\, \leftrightarrow \ket{\text{D}_{5/2}, m_J=3/2}$ transition with a $729\,$nm laser pulse. At the end, the quantum state is read out for all individual ions via spatially resolved fluorescence detection at $397\,$nm recorded with the EMCCD camera. In this work, the ions is outside the LDR such that we cannot extract the phonon number from the sidebands spectra since the sizable asymmetry cannot be observed even after PGC. Therefore, the final phonon number is determined by fitting the Rabi oscillations of the carrier and first order blue and red sidebands for the single ion, and in the same way for the  multi-ion crystals.


\section{Experimental characterization of PGC} \label{experimantal characterization PGC}

\begin{figure}[t!]
\centering
\includegraphics[width=0.7\textwidth]{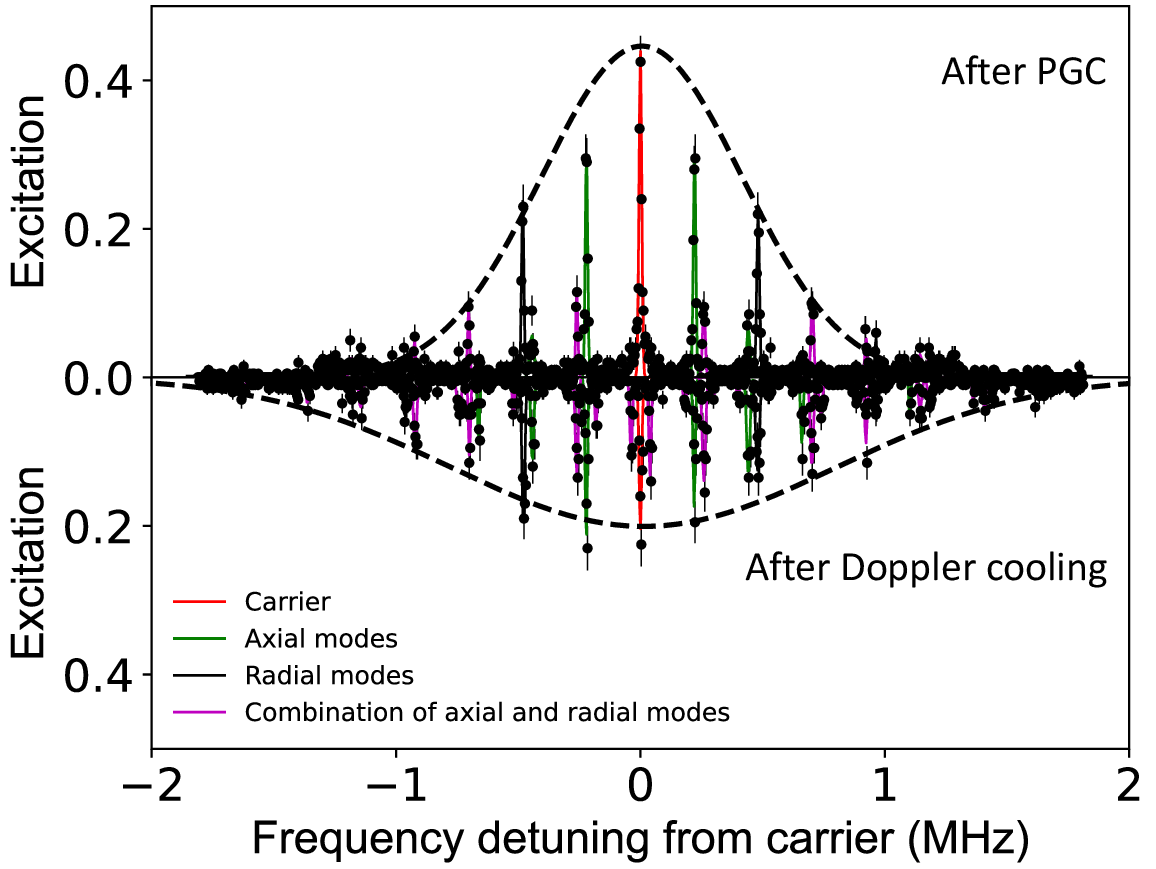}
\caption{Spectra of motional states. The $\ket{\text{S}_{1/2}, m_J=1/2}\,\leftrightarrow\,\ket{\text{D}_{5/2}, m_J = 3/2 }$ transition is probed with a laser pulse at $729\,$nm after PGC (top) and after Doppler cooling (bottom). Each experimental point (black solid circles) is the average of 200 measurements. The envelope fit (dashed black lines) include all the sidebands by taking  into account the maxima of Gaussian fits to the carrier transition (red), axial sidebands (green), radial sidebands (black solid line) and combinations of axial and radial sidebands (magenta). } 
\label{fig:SpecWithsame729powerGaussionAll}
\end{figure}

Three dimensional PGC is compared to Doppler cooling on a single Ca$^+$ ion by sideband spectroscopy of the $\ket{\text{S}_{1/2}, m_J=1/2}\,\leftrightarrow \ket{\text{D}_{5/2}, m_J = 3/2}$ transition, as shown in Fig.\,\ref{fig:SpecWithsame729powerGaussionAll}. In this measurement, the $393\,$nm laser light is red-detuned by $2\pi\times180\,$MHz, as determined with a wavelength meter\footnote{WS7, HighFinesse}. The intensity of the $729\,$nm ligth for both measurements is same. The sideband data are fitted with individual Gaussian profiles. The higher-order sidebands involving the axial and radial motional modes and the higher order sidebands are considerably suppressed for the PGC case. Furthermore, a Gaussian envelope fit to the sideband amplitudes reveals a linewidth for the PGC case reduced by a factor of $1.92(6)$ compared to that with Doppler cooling.

\begin{figure}[t!]
\centering
\includegraphics[width=0.7\textwidth]{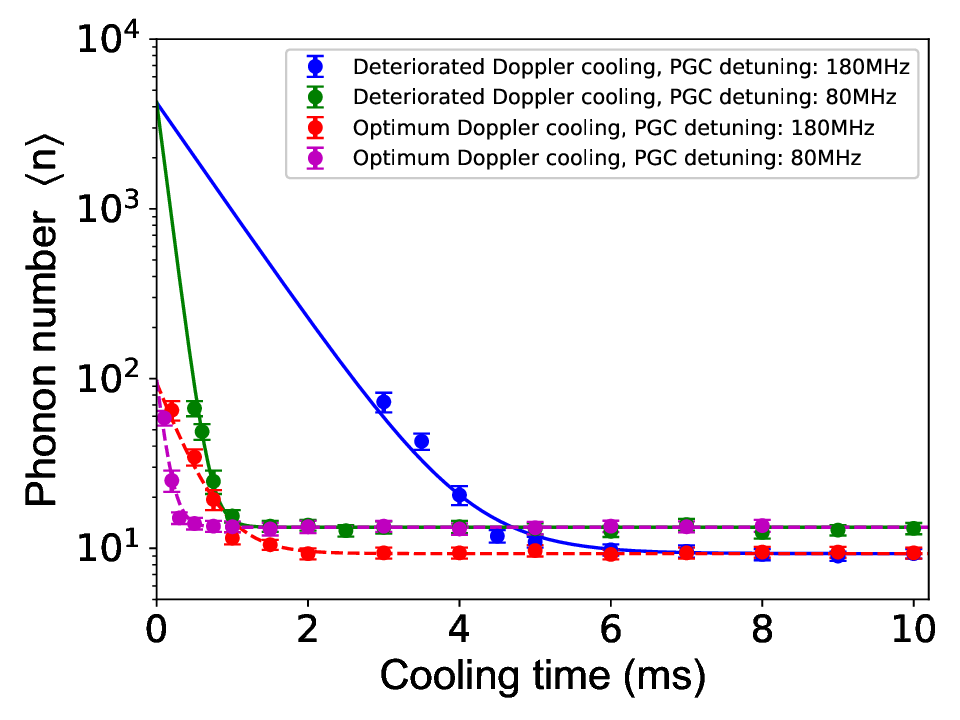}
\caption{PGC performance for different cooling times and different initial phonon numbers. The initial state is prepared by Doppler cooling, which is either optimized or deliberately spoiled by tuning the Doppler cooling light close to resonance ($\Delta \ll \Gamma$). The characteristic cooling times are $\lesssim 7.0\,$ms\,(blue circles fit with dashed line) and  $\lesssim 1.5\,$ms\,(green) and the final mean phonon numbers 9.3(7) and 13.3(11), for PGC-light detuning of $2\pi\times180\,$MHz and $2\pi\times80\,$MHz, respectively. With low initial phonon numbers, the cooling times are decreased to $\lesssim 3.0\,$ms (red) and $\lesssim 0.75\,$ms (magenta), respectively.
The data in this figure were taken without the cavity filter.}
\label{fig:CoolingtimePreDopplerVSdetuningN.eps}
\end{figure}

To investigate the robustness of the presented PGC method, we initialize the ion with high initial phonon number occupation($>4000$ phonons, extracted from the extrapolated logarithmic fit in Fig.\,\ref{fig:CoolingtimePreDopplerVSdetuningN.eps}) by artificially deteriorating Doppler cooling via detuning the $397\,$nm light close to resonance ($\Delta\ll\Gamma$) of the S$_{1/2}$ to P$_{1/2}$ transition. The cooling results compared with PGC after optimal Doppler cooling are shown in Fig.\,\ref{fig:CoolingtimePreDopplerVSdetuningN.eps}.  With sufficient cooling time, identical final mean phonon numbers can be achieved independent of the initial ion temperature. For a detuning of $2\pi\times180\,$MHz, the equilibrium cooling time is more than twice as long for high initial phonon numbers as for low initial phonon numbers. The PGC time can be reduced by decreasing the detuning of PGC light to $2\pi\times80\,$MHz, which in turn increases the scattering rate of PGC light. We have also realized PGC for detunings as small as $30\,$MHz, and this was decreasing the cooling time to $\leq$35$\,\mu$s at the price of an elevated mean phonon number. With this small detuning of 30 MHz, the cooling time is decreased about 2 orders of magnitude compared with the optimum configuration shown in Fig.\,\ref{fig:CoolingtimePreDopplerVSdetuningN.eps}. However, the final phonon numbers are just about half Doppler cooing limit. For red-detuned PGC, we can explore the cross-over to Doppler cooling.

\begin{figure}[t!]
\centering
\includegraphics[width=0.7\textwidth]{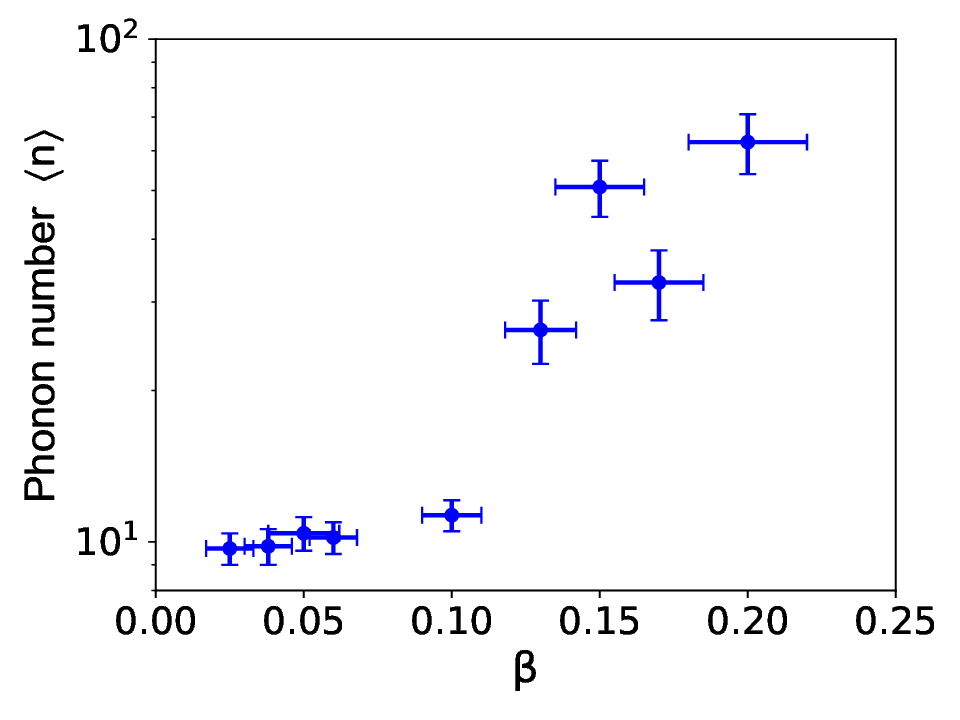}
\caption{Mean phonon numbers with the transition coupling strength on the micromotion (MM) sideband. The residual micromoition is changed by scanning the compensation voltages in one of the radial directions.} The performance of PGC drops significantly for a modulation index $\beta > 0.1$.
\label{fig:MMComVolLog.eps}
\end{figure}

For three-dimensional PGC, residual micromotion as one of the crucial heating sources for trapped ions is detrimental, especially for weak radial trap potentials. Therefore, we also investigate the influence of micromotion on PGC by adjusting micromotion-compensation voltages. The micromotion modulation index $\beta$ is determined from the ratio of the excitation strength at the carrier and the micromotion-sideband frequency.
We find PGC to be robust against residual micromotion corresponding to a range where the micromotion-compensation DC voltage is changed by $1.6\,$V (corresponding to a displacement of 560(35)\,nm from the trap center) and thus a micromotion-modulation index $\beta\leq 0.1$, see Fig.\,\ref{fig:MMComVolLog.eps}. However, micromotion is a decisive factor for three dimensional PGC. Therefore, the residual micromotion has to be compensated within a proper range. Otherwise, the ions will be heated up during the cooling time duration of PGC.

\begin{figure}[t!]
\centering
\includegraphics[width=0.7\textwidth]{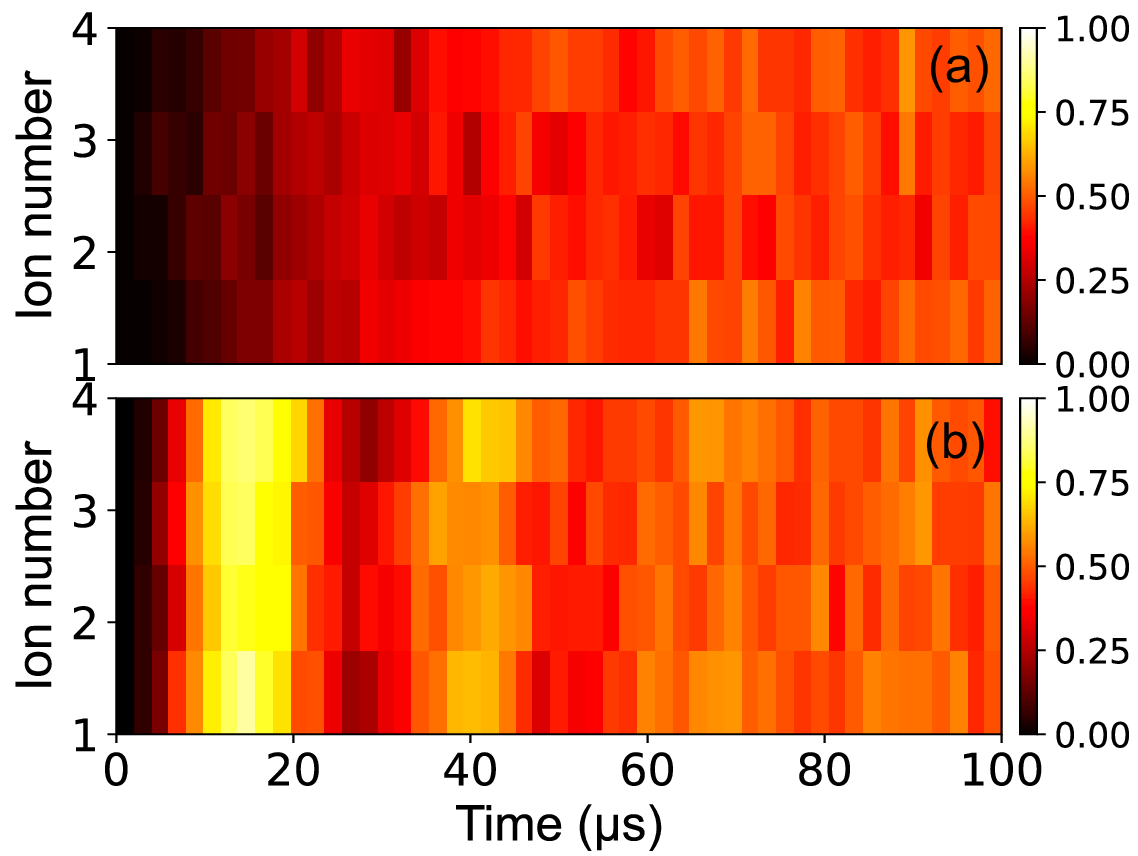}
\caption{Rabi dynamics for a 4-ion linear crystal recorded individually on the carrier transition with an axial trap frequency of $190\,$kHz. For Doppler cooled ions~(a), no oscillations are visible, while they can clearly be observed after PGC~(b).}
\label{fig:4IonsRaibi.eps}
\end{figure}

We extend PGC to a 4-ion linear crystal confined in a trap potential, where the radial trap frequencies degeneracy is lifted by applying $3.0\,$V on the DC electrode pair. The trap frequencies are $(\omega_x, \omega_y, \omega_z) = 2\pi\times (427, 505, 190)\,$kHz. The optical parameters used here are the same as the optimized for a single ion. We determine the axial phonon number by measuring Rabi oscillations on the S$_{1/2}\leftrightarrow$D$_{5/2}$ transition. The Rabi oscillations of all four ions on the carrier transition are recorded individually to demonstrate the improved cooling of PGC with respect to Doppler cooling as shown in Fig.\,\ref{fig:4IonsRaibi.eps}. For the Doppler-cooled ion crystal, no Rabi oscillations are visible for low trap frequencies, as shown in Fig.\,\ref{fig:4IonsRaibi.eps}\,(a). This is due to the averaging of Rabi oscillations for different phonon numbers, coined spectator modes effect\,\cite{Wineland1998}. However, for the PGC case, Rabi oscillations can be explicitly observed, see Fig.\,\ref{fig:4IonsRaibi.eps}\,(b). Since all motional modes contribute to the Rabi dynamics of the carrier transition, this result indicates a low phonon occupation number for each motional mode of the 4-ion linear crystal. In contrast to the previous work\,\cite{Joshi2020} our experimental geometry allows for addressing all 3-D motional modes. Having a projection on the radial modes also allows for studying the PG cooling of all modes, see Fig.\ref{fig:SpecWithsame729powerGaussionAll}. Also, the effect of a micromotion compensation on PGC can be studied in this geometry, because radial micromotion is typically much larger as compared to the axial one.

\begin{figure}[t!]
\centering
\includegraphics[width=0.75\textwidth]{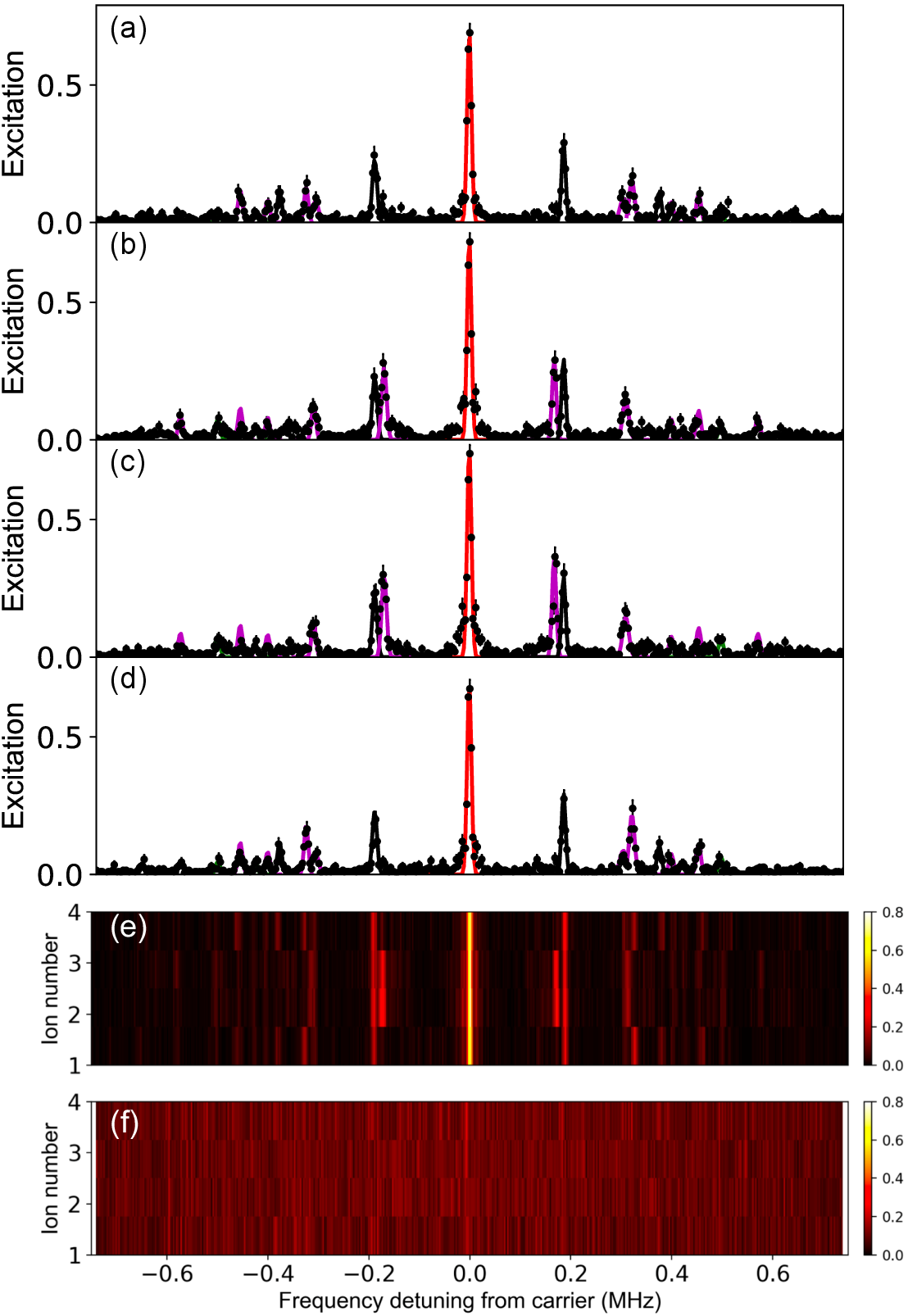}\llap{\makebox[\wd2][r]{\raisebox{16.32cm}
{\includegraphics[height=1.6cm]{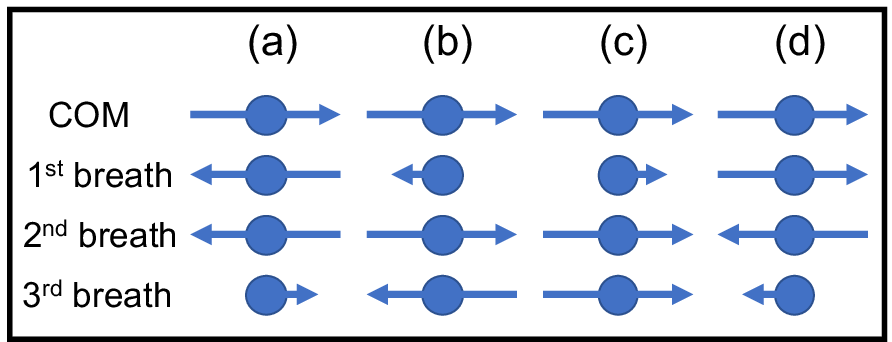}}}}
\caption{Sideband spectra of a 4-ion linear crystal after PGC recorded on individual ions display carrier and sideband, depending on the site in the crystal ((a)-(d)), see inset with eigenvectors of different length for the axial modes. We fit Gaussian profiles to the data: carrier (red line), and axial modes (black lines), and the mixing between the radial COM and the first breathing mode at 505kHz-330kHz=175kHz (magenta). Note, that this resonance only appears for central ions (a) and (d) with a strong first breathing eigenvector. In contrast, the fourth axial mode at 582kHz is only excited sufficiently strong on the innermost ions (b) and (c). Each experimental point (black solid circles) is the average of 200 measurements. Note, that for the central ions in the crystal, the first breathing mode is suppressed, while the fourth mode is suppressed for the outer ions. In the density plot the difference between PGC (e) and Doppler cooling (f) is clarified, as in the latter case sidebands are washed out.}
\label{fig:4IonsSpecInd.eps}
\end{figure}

The individual motional modes of the 4-ion linear crystal are investigated with spatially resolved sideband spectroscopy, as shown in Fig.\,\ref{fig:4IonsSpecInd.eps}. All the sidebands including center-of-mass (COM) and breathing modes of four ions can be distinctly resolved after PGC as shown in Fig.\,\ref{fig:4IonsSpecInd.eps}\,(a)-(e). In contrast, for the Doppler cooled case shown in Fig.\,\ref{fig:4IonsSpecInd.eps}\,(f) the high phonon numbers for each motional mode, and therefore the emergence of higher harmonics, prevent the resolution of individual sidebands. The final mean phonon number of the axial mode is evaluated as $12.5(7)$ after PGC by measuring Rabi oscillations of the carrier transition and the first axial sidebands of the COM mode of the inner ions.

\section{Technique for improving PGC efficiency}\label{techniquee for improving PGC}

An external-cavity diode laser is used for PGC in this work. In this case, resonant scattering events due to the amplified spontaneous emission (ASE) of the laser diode represent an additional heating source. The final temperature after PGC would be limited by this effect, especially if the frequency detuning is increased in order to approach the theoretical temperature limit~\cite{Dalibard1989,Joshi2020}. For large detunings, the lower cooling rate will be surpassed by the heating rate of the ASE background.
To investigate this regime, we build a Fabry-P\'{e}rot cavity to purify the laser spectrum. Two mirrors with reflectivity of $ 99(1)\,\% $ are spaced by $8\,$cm to form an optical resonator with a free spectral range (FSR) of $1.87\,$GHz and a finesse of about $\mathcal{F}=310$. The measured extinction of ASE is in good agreement with the expected value of $(1-R)^2\approx10^{-4}$. To stabilize the cavity and  power of the transmitted light, the cavity is frequency locked at low Fourier frequencies to the 393 nm laser by a Piezo-transducer (PZT). The 393 nm laser is frequency locked at high Fourier frequencies to the cavity by the laser current which makes the laser frequency can follow the cavity quickly. At the same time, the 393 nm laser frequency is kept stable using feedback from the high precision wavelength meter WS07. The resulting cavity filter with linewidth $\Delta\nu\approx6\,$MHz, consequently, is able to sufficiently suppress spectral impurities of the laser light.  

\begin{figure}[t!]
\includegraphics[width=0.7\textwidth]{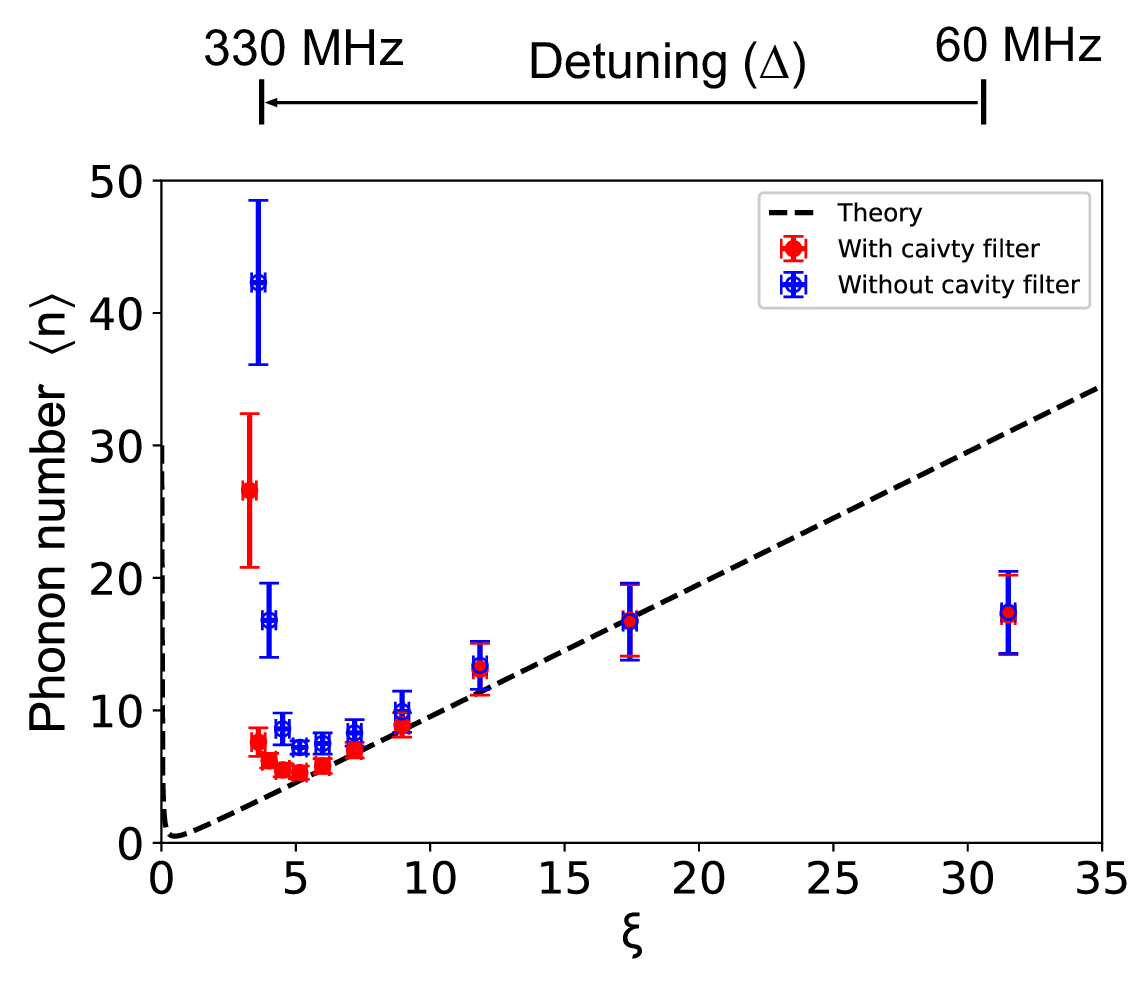}
\centering
\caption{Mean phonon number after PGC for different $\xi= s\,\Delta/(3\omega_z)$, realized by changing the detuning $\Delta$ with (red solid circles) and without (blue hollow circles) spectral purification via the cavity filter, respectively. The minimum mean phonon number at a detuning of $2\pi \times210\,$MHz is reduced by $2.0(3)$ phonons. The dashed line indicates the theoretical predicted mean phonon number with $\xi$.}
\label{fig:PhononVSdetuningLog285Theory.eps}
\end{figure}

In an ideal case, the final mean phonon number is expected to be $\langle n_0(\xi)\rangle=\xi+1/(4\xi)-1/2 $ for a single ion~\cite{Joshi2020}, where $\xi= s\,\Delta/(3\omega_z)$ is a dimensionless parameter, $\Delta$ is the frequency detuning of $393\,$nm light, and $\omega_z$ is the trap frequency in the axial direction. The saturation parameter  is defined as $s = \frac{\Omega^2/2}{ \Gamma^2/4+\Delta^2}$, with $\Gamma=2\pi\times23\,$MHz the width of the excited state P$_{3/2}$ and $\Omega$ the Rabi frequency. To investigate how well we approach this limit using light with purified spectrum  we measure the final mean phonon number for different frequency detunings of the $393\,$nm light with fixed power. For this experiment, the trap frequencies are $(\omega_x, \omega_y, \omega_z) = 2\pi\times(483, 480, 285)\,$kHz. Figure\,\ref{fig:PhononVSdetuningLog285Theory.eps} shows the results with (red) and without (blue) cavity filter, respectively, and the theory fit with the laser intensity as the free parameter. Suggests that theoretical estimation a minimum mean phonon number of $1/2$ can be achieved for $\xi=1/2$ when the optical well depth~($2s\,\Delta/3$) equals the trap frequency $\omega_z$. From our measurement, the phonon number is in agreement with the theory prediction for $\xi>5$. However, the minimum phonon number $1/2$ predicted by the theory cannot be reached in our experiment since the trap potential is weak. Indeed, when the detuning is decreased to reduce the optical potential closer to the theoretical optimum, the optical pumping rate also decreases and, eventually, PGC cannot counteract the trap heating rate. At the frequency detuning of $2\pi\times210\,$MHz, the minimal achieved phonon number is reduced  by $2.0(3)$ when using the cavity filter.

\begin{figure}[t!]
\includegraphics[width=0.7\textwidth]{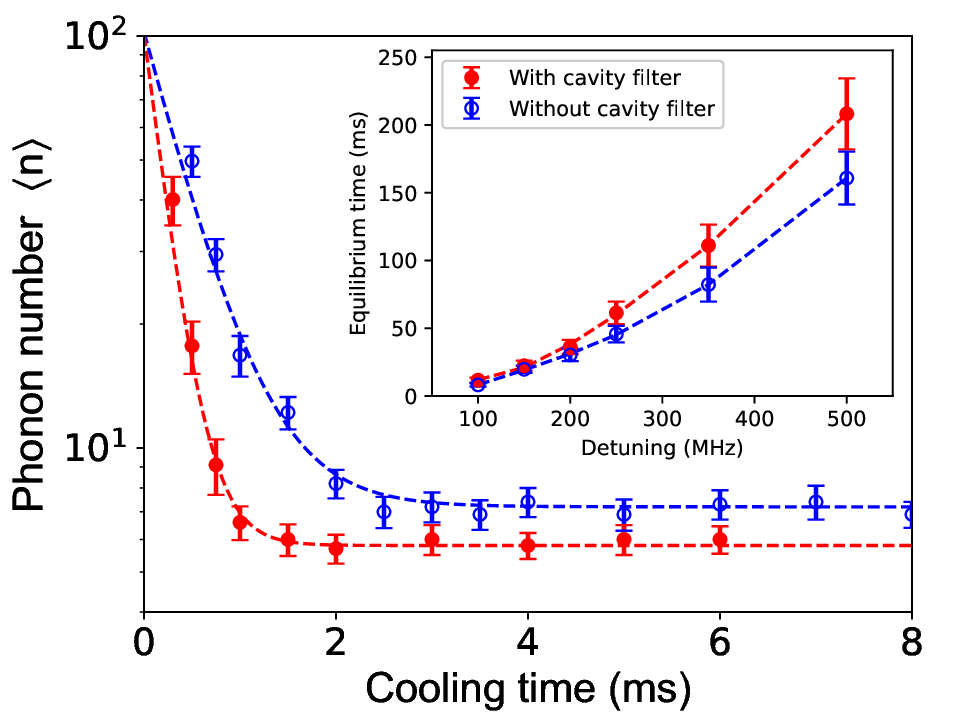}
\centering
\caption{Cooling dynamics of PGC with (red solid circles) and without (blue hollow circles) cavity filter for fixed detuning of $2\pi\times210\,$MHz and fixed intensity of the laser beams. The equilibrium cooling rate is improved by a factor of 2.1(2) when the cavity filter is used. Inset: pumping dynamics from one S$_{1/2}$ Zeeman state to the equilibrium by the $393\,$nm light beams with (red solid circles) and without (blue hollow circles) cavity filter, respectively. The increased pumping rate without the cavity filter is caused by additional resonant scattering events due to the amplified spontaneous emission background.}
\label{fig:CoolingTimeWithInset285.eps}
\end{figure}

To further investigate the effect of the ASE, we measure the cooling dynamics of PGC with and without cavity filter with a fixed frequency detuning of $2\pi\times210\,$MHz and fixed intensity, see Fig.\,~\ref{fig:CoolingTimeWithInset285.eps}. The observed cooling time is improved by a factor of 2.1(2). To investigate the resonant scattering rate caused by the ASE we measured the pumping time from the $m_J = +1/2$ Zeeman state to the equilibrium distribution in the ground S$_{1/2}$ state, as follows. After Doppler cooling, the ion is prepared in the $\ket{\text{S}_{1/2}, m_J = +1/2}$ ground state via optical pumping with $729\,$nm and $854\,$ nm laser light. Afterwards, the $393\,$nm laser is applied to the ion with a varying duration resulting in partial population of the $\ket{\text{S}_{1/2}, m_J = -1/2}$ level. Finally, the population in the $\ket{\text{S}_{1/2}, m_J = +1/2}$ state is shelved to the D$_{5/2}$ level with a $729\,$nm laser pulse and the residual population in the $\ket{\text{S}_{1/2}, m_J = +1/2}$ state is detected by a fluorescence measurement. By scanning the pulse duration of the $393\,$nm laser light, the time to reach an equilibrium between the Zeeman ground states is measured for different frequency detunings as shown in the inset of Fig.\,\ref{fig:CoolingTimeWithInset285.eps}. The increased equilibriation time when the cavity filter is used compared with the case without the cavity filter is a result of reduced ASE background of the diode laser leading to a decreased resonant scattering rate. The suppression of the ASE background of the diode laser by the Fabry-Pérot cavity filter results in a reduced additional heating rate and, therefore, an improved cooling rate and final mean phonon number after PGC. The cavity-filter technique provides an option to improve PGC, even with diode lasers with excessive ASE.

\begin{figure}[t!]
\includegraphics[width=0.7\textwidth]{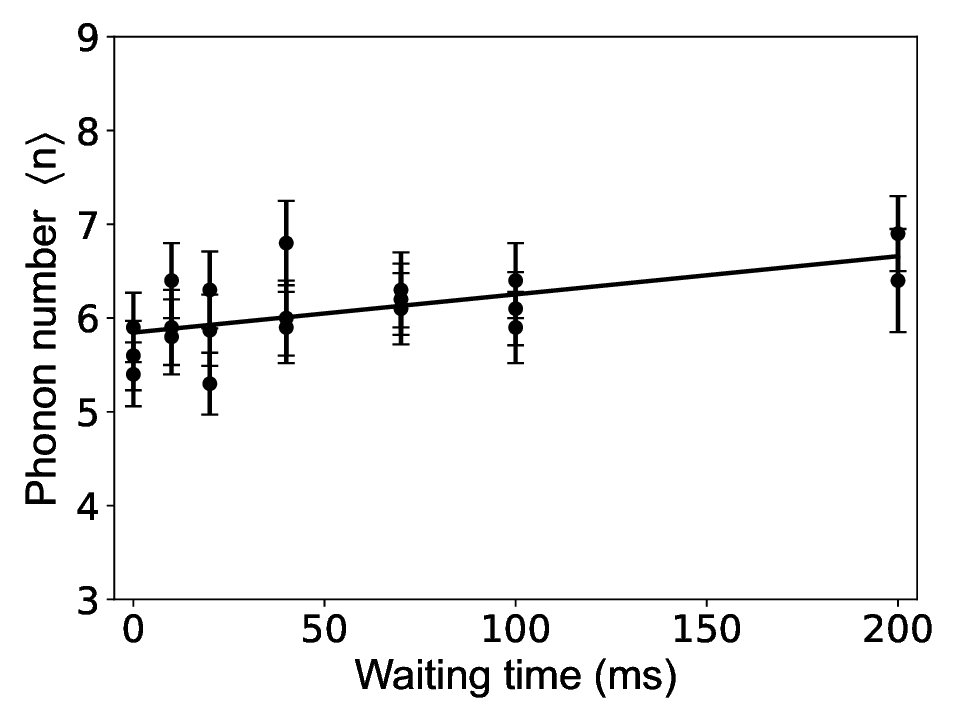}
\centering
\caption{Heating rate of axial motional mode. The mean phonon number is measured for different waiting times after PGC. A linear fit reveals a heating rate of $4.1(8)$ phonons per second for a trap frequency of $2\pi\times285\,$kHz.}
\label{fig:HeatingRate285}
\end{figure}

To test whether the ultimate PGC performance would be limited by the heating rate of the trap, we evaluate heating in the axial motional mode. The final mean phonon number is measured for different waiting times after PGC. A linear fit to the data in Fig.\,\ref{fig:HeatingRate285} reveals a heating rate of $4.1(8)$ phonons per second at an axial trap frequency of $2\pi\times285\,$kHz. This rate is of the same order of magnitude as the cooling rate close to final temperatures at the detuning of $2\pi\times210\,$MHz and contributes to the limitation in achievable final mean phonon number of PGC.

\section{Conclusion and outlook} \label{conclusion and outlook}

We implement three-dimensional polarization gradient cooling for trapped Ca$^+$ ions in a linear Paul trap with initial temperatures corresponding to ion motion being far outside the Lamb-Dicke regime. Sideband spectroscopy after PGC shows all motional modes to be well below the Doppler-cooling limit.  Efficient cooling close to the ground state even with elevated initial phonon numbers ($>4000$ phonons) demonstrates the robustness of the presented method. The deterministic influence of residual micromotion on PGC is discussed. While our results fit well to a rudimentary semiclassical PGC theory, the quantitative comparison in the crossover regime between PGC and Doppler of ion crystals might need a more refined theory development.

We have shown PGC for single ions and extended the cooling technique to linear crystals consisting of four ions reaching a final mean phonon number for the common mode of $12.5(7)$ at common mode of the axial trap frequency of $2\pi\times190\,$kHz. Furthermore, we find that the spectral impurity of the laser field adversely affects the cooling rate and cooling limit of PGC. Thus, we employ a Fabry-Pérot cavity to suppress amplified spontaneous emission of the diode laser. The mean phonon number is improved by $2.0(3)$ phonons and the cooling time is decreased by a factor of $2.1(2)$. A mean phonon number limit of $5.4(4)$ is achieved at a trap frequency of $2\pi\times285\,$kHz in the axial direction. The heating rate of the trap is estimated to be $4.1(8)$ phonons per second, which, in combination with the estimates of the cooling rate presented in Sec.\,\ref{techniquee for improving PGC}, indicates that the lowest achievable phonon number is several phonons per mode. 

The cavity-filter technique provides an option to improve PGC, even with diode lasers with an ASE component in the emission spectrum. This robust PGC scheme will be extended to larger ion crystals, including 3D formations, and can be beneficial for capturing and sympathetically cooling injected ``impurity'' ions, e.g. thorium \cite{Berning2019}, below the Doppler cooling limit.


\section{Acknowledgements}

WL thanks M. K. Joshi for helpful discussions. We  thank Ulrich Poschinger and Daniel Wessel for careful reading and helpful comments. This work was supported in part by the Cluster of Excellence ``Precision Physics, Fundamental Interactions, Structure of Matter'' (PRISMA+ EXC 2118/1) and Helmholtz Excellence Network ExNet020, funded by the German Research Foundation (DFG) within the German Excellence Strategy (Project ID 39083149), the DFG Reinhart Koselleck project and the Deutsche Forschungsgemeinschaft (DFG, German Research Foundation) – Project-ID 429529648 – TRR 306 QuCoLiMa (“Quantum Cooperativity of Light and Matter”). WL thanks for the financial support by the “China-Germany Postdoctoral Exchange Program”.  


\end{document}